\documentclass[hyper,12pt,paper]{article}
\pdfoutput=1
\usepackage{jheppub}

\usepackage{ccaption}
\usepackage{graphicx,amsmath,amssymb,multirow,array,bm,mathrsfs}
\usepackage{dsfont}
\newenvironment{definition}[1][Definition]{\begin{trivlist}
\item[\hskip \labelsep {\bfseries #1}]}{\end{trivlist}}

\newenvironment{theorem}[1][Theorem:]{\begin{trivlist}
\item[\hskip \labelsep {\bfseries #1}]}{\end{trivlist}}

  \captionnamefont{\bfseries}
  \captiontitlefont{\small\sffamily}
  \captiondelim{: }
  \hangcaption

\title{Emergent Gravity requires (kinematic) non-locality}

\author[]{Donald Marolf,}

\affiliation{Department of Physics, University of California, Santa Barbara, CA 93106, USA.}
\emailAdd{marolf@physics.ucsb.edu}

\abstract{This work refines arguments forbidding non-linear dynamical gravity from appearing in the low energy effective description of field theories with local kinematics, even for those with instantaneous long-range interactions.  Specifically, we note that gravitational theories with universal coupling to energy -- an intrinsically non-linear phenomenon -- are characterized by Hamiltonians that are pure boundary terms on shell.  In order for this to be the low energy effective description of a field theory with local kinematics, all bulk dynamics must be frozen and thus irrelevant to the construction.  The result applies to theories defined either on a lattice or in the continuum, and requires neither Lorentz-invariance nor translation-invariance.
}

\begin{document}
	
\maketitle
\flushbottom

%~~~~~~~~~~~~~~~~~~~~~~~~~~~~~~~~~~~~~~~~~~~~~~~
\section{Introduction}
\label{sec:intro}
%~~~~~~~~~~~~~~~~~~~~~~~~~~~~~~~~~~~~~~~~~~~~~~

Attempts to directly quantize the gravitational field encounter well-known difficulties associated with lack of perturbative renormalizeability, the black hole information problem, and the lack of local observables due to invariance under diffeomorphism gauge symmetry (which, from the active point of view, moves spacetime points from one location to another).  While it remains possible that any or all of these issues may one day be surmounted, it is nevertheless interesting to ask whether diffeomorphism-invariant gravity might emerge an effective approximate description of a system that is inherently better behaved at the microscopic level.

Most leading approaches to quantum gravity embody ideas along these lines. String theory, loop quantum gravity, causal sets \cite{Bombelli:1987aa}, and causal dynamical triangulations \cite{Ambjorn:1998xu} (see e.g. \cite{Ambjorn:2013apa} for a recent overview) all propose that smooth classical geometries arise only in appropriate semiclassical limits.\footnote{Asymptotic safety is the most prominent exception; see e.g. \cite{Reuter:2012id,Nagy:2012ef} for recent reviews.}  But the structures underlying these theories again involve novel physics that is difficult to control.  So it is natural to ask if gravity can arise from more familiar systems such as field theories with local kinematics.   Examples of such proposals include \cite{Akama1978,Amati:1981rf,Floreanini:1990cf,RandjbarDaemi:1994ng,Volovik:2000ua,2001Sci...294..823Z,Wetterich:2003wr,Hu:2005ub,Barcelo:2005fc,Xu2006cond.mat..2443X,Xu2006PhRvB..74v4433X,Weinfurtner:2007br,Volovik:2008dd,Gu:2009jh,Xu:2010eg,2012AnPhy.327.1146Z,Alfaro:2012fs}. Below, we argue that such scenarios can succeed only if the map to gravitational degrees of freedom involves long-range non-locality; i.e., only if the notions of locality are very different in the two descriptions.

As has been well-known for some time, the (spin-2) Weinberg-Witten theorem \cite{Weinberg:1980kq} already excludes the emergence of gravity from local Poincar\'e-invariant field theories.  In particular, it forbids such theories from containing an interacting massless spin-2 degree of freedom in its spectrum of asymptotic states.  While clear and concise, the technical assumption of Poincar\'e-invariance appears to leave open many doors for exploration.  For example, one might attempt to evade the theorem by working on a lattice as in e.g. \cite{RandjbarDaemi:1994ng,Xu2006cond.mat..2443X,Xu2006PhRvB..74v4433X,Gu:2009jh,Xu:2010eg}, or by using other structures that break this symmetry.

However, as noted in e.g. \cite{WittenTalk,Carlip:2012wa}, the lack of local observables in quantum gravity suggests a more general result forbidding diffeomorphism-invariant gravity arising as the effective description of {\it any} theory with sufficiently interesting local observables.  Our purpose here is to make this precise.  Since any theory can be made diffeomorphism-invariant via a process known as parametrization (see e.g. \cite{Arnowitt:1959ah,Deser:1960zzc2,Arnowitt:1960es,Arnowitt:1960zzc,Arnowitt:1961zz,Arnowitt:1962hi,Deser:1960zzc2,Kuchar:1976yw,Kuchar:1976yx,Kuchar:1976yy,Kuchar:1977}), we follow \cite{Andrade:2010hx} in using the gravitational Gauss law to distinguish theories with sufficiently `interesting' diffeomorphism-invariance.

Before stating our technical result in section \ref{def} below, let us therefore take a moment to explain this idea in broadly accessible terms.  We first recall that (non-relativistic) Newtonian gravity can be formulated in terms of a gravitational potential $\phi$ that satisfies a Poisson equation $\nabla \phi = 4 \pi G \rho_M$ sourced by the mass-density $\rho_M$.  As a result, the total mass $M = \int_V \rho_M \ dV$ inside a volume $V$ (with volume element $dV$) can be expressed as the boundary term $M =  \frac{1}{4\pi G} \int_{\partial V} dS \ n^i \partial_i \phi$ where $\partial_i = \frac{\partial}{\partial x^i}$ denotes derivatives with respect to spatial coordinates $x^i$, $n^i$ is the unit (outward-pointing) normal to the boundary $\partial V$, and $dS$ is the area element on $\partial V$.  This is just the Newtonian gravity analogue of Gauss' Law from electrostatics. Now, in relativistic theories, the gravitational field couples not just to mass, but to all forms of energy.  As a result, in the presence of appropriate boundary conditions one finds a corresponding Gauss-law-like boundary integral that encodes the total energy $E$; see e.g. \cite{Handbook} for a recent review.  This Gauss-law for energy will turn out to be the critical feature that forbids the theory from arising as an effective description; the full Lorentz-invariance that originally motivated the coupling to energy is not required\footnote{Lorentz-violating theories that couple universally to energy may be constructed in analogy with Ho\v{r}ava-Lifhshitz gravity \cite{Horava:2009uw}, interchanging the roles of space and time and replacing the extrinsic curvature of a preferred foliation with the proper acceleration of a preferred family of worldlines.}.  It is useful to mention here that the Gauss-law property is inherently non-linear due to the fact that the energy source term also receives contributions from the gravitational field.  As a result, in parallel with the Weinberg-Witten theorem \cite{Weinberg:1980kq}, our arguments place no constraints on the emergence of strictly linear spin-2 degrees of freedom.

As we explain below,
the Gauss-law property will imply that gravity can be a good effective description of a theory with local kinematics\footnote{The dynamics is required to be local in time and generated by a Hamiltonian.  However, the Hamiltonian can be non-local in space.  We explicitly allow instantaneous long-range interactions.} only in limits where the bulk dynamics freezes out away from the boundary. While there is nothing wrong with such freeze out in an of itself\footnote{There are of course many theories where bulk excitations are gapped at low energy.}, an interesting effective gravitational description should remain non-trivial in the bulk of the spacetime.  Consistency then requires that bulk gravitational physics be the effective description of purely boundary dynamics in the original theory.  Modulo anomalies, the original bulk theory served no purpose in the construction and may be discarded.  As a result, the notions of bulk vs. boundary are completely different in the original kinematically-local theory and the effective gravitational description.  This is the requisite non-locality referred to in the title.  The reader will note that it also describes a paradigm embodied in string theory by gauge/gravity duality (e.g. \cite{Banks:1996vh,Maldacena:1997re}).

\section{Definitions and Results}
\label{def}

We begin the technical treatment by making two definitions that will allow us to sharply state our result.  Each definition is followed by comments to provide clarity.  Discussion of the main result will appear in section \ref{disc}.

\begin{definition}
{\bf I.} A gravitational theory with {\it universal coupling to energy} is one for which, in the presence of any boundary conditions for which a Hamiltonian exists, the total energy can be written as the integral over the boundary of space at each time of some local function of the gravitational field and its derivatives.  Below, we refer to the integrand of this boundary integral as the gravitational flux.  We require the gravitational flux to be an observable (i.e., it is invariant under gauge transformations allowed by the given boundary conditions).
\end{definition}
We now make several remarks to clarify this definition. See e.g. \cite{Handbook} for any definitions and for further discussion of the examples below.  We will use the term Riemann-curvature gravity theories to refer to Einstein-Hilbert gravity together together with its higher-derivative generalizations described by Lagrangians that are local scalar functions of the Riemann tensor and its derivatives.

\begin{enumerate}

\item{} {\bf Simultaneity:}  The phrasing implies that the spacetime boundary admits a notion of ``each time;" i.e., of which points on the boundary are simultaneous.  This notion need not be unique; e.g., for Riemann-curvature gravity with either anti-de Sitter boundary conditions or Dirichlet boundary conditions at a finite wall, any time-function on the boundary may be used to define simultaneity so long as all pairs of points on its level surfaces are spacelike separated.   The notion of simultaneity is also allowed to be trivial as in asymptotically flat Riemann-curvature gravity where the boundary of space should be interpreted as spacelike infinity ($i^0$) and, in the usual representation, all points at spacelike infinity are simultaneous.  Indeed, the entire notion of ``boundary of space" can be trivial so long as the total energy vanishes identically in such cases; Riemann-curvature gravity for closed cosmologies provides an example.

\item{} {\bf Total Energy:} This quantity is defined to be the generator of (asymptotic) time-evolution; i.e., it is the Hamiltonian.  This time evolution need not be a symmetry, so the Hamiltonian may have explicit time-dependence.

\item{} {\bf Observable Gravitational Flux:} We remind the reader that the gravitational flux at the boundary is indeed gauge-invariant in Riemann-curvature gravity since it can be defined as the variation of the action (see e.g. \cite{Brown:1992br,Witten:1998qj,Henningson:1998gx,Balasubramanian:1999re,Mann:2005yr}) with respect to boundary conditions (which are by definition gauge-invariant).  It may also useful to mention that, while often not presented in this form, in Einstein-Hilbert gravity with asymptotically flat \cite{Ashtekar:1978zz} or asymptotically AdS boundary conditions \cite{Ashtekar:1984zz,Ashtekar:1999jx} the gravitational flux may be written in terms of the Weyl tensor at the boundary.  In this form it more closely resembles the familiar electric flux computed from the field strength of a vector gauge field.

\item{} {\bf Heuristics, examples, and contrasting theories:} The idea behind calling the above property ``universal coupling to energy" is that there is an aspect of the gravitational field (namely the above boundary integral) which directly gives the total energy of the system.  Any Riemann-curvature theory satisfies this definition (see e.g. \cite{Iyer:1994ys}).  In contrast, Ho\v{r}ava-Lifshitz gravity  \cite{Horava:2009uw} and massive gravity theories (see e.g. \cite{Hinterbichler:2011tt} for a recent review) do not have universal coupling to energy in this sense. We will not discuss such theories further except that to note that their behavior is generally rather distinct \cite{Henneaux:2009zb,Deser:2013rxa,Deser:2013qza} from theories with universal coupling.
\end{enumerate}

\begin{definition}
{\bf II.} Whether defined in the continuum or on a spatial or space-time lattice, a theory will be said to be {\it kinematically local} iff the commutator of two gauge-invariant local Heisenberg-picture operators (with at least one bosonic) vanishes when evaluated at different spatial locations at a common time.   We also assume that time-evolution is generated by some Hamiltonian.
\end{definition}
We again provide clarifying comments below

\begin{enumerate}
\item{} {\bf Simultaneity:} We require the theory to have a concept of bulk simultaneity (i.e., when two spacetime events occur at the same time).  We assume this to be a background structure independent of dynamical fields.   As above, this notion need not be unique; i.e., in a relativistic theory it will suffice to choose any time-function that is constant on spacelike surfaces\footnote{By which we mean that there is no causal connection between any two points on the surface.}.  So any theory built in the usual local way from scalar, spinor, or vector fields in local in this sense.

\item{} {\bf Heisenberg Picture:}  We assume the existence of a Heisenberg picture, in which gauge-invariant operators at each position $\vec x$ satisfy $- i  \hbar \frac{\partial}{\partial t} {\cal O}(\vec x, t) = [H(t), {\cal O}(\vec x, t)]$ for some (perhaps time-dependent) Hamiltonian $H(t)$.  In this sense the dynamics is local in time, though $H(t)$ may be arbitrarily non-local in space.  In particular, instantaneous long-range interactions are allowed.

\item{} {\bf Bosonic Operators:}  It is sufficient for our purposes to define gauge-invariant operators to be bosonic when they commute with all local gauge-invariant operators located at different positions in space at the same time.  While we referred to local operators above, in a lattice theory it is natural to also consider bosonic operators $B$ built from multiple nearby lattice sites; e.g., the product of two free fermions at adjacent sites may be considered a local bosonic operator.  In that case we require it to commute with all operators whose support does not contain the lattice sites from which our operator $B$ was built.
\end{enumerate}

Combining the above definitions leads quickly to the desired result.  We begin by assuming the theory with local kinematics to admit some limit where it is effectively described by a gravitational theory with universal coupling to energy.  We take this to mean that notions of time-evolution embodied by the above two definitions coincide.  The time evolution of the local theory is then generated by a Hamiltonian, which by definition I can be written as an integral over the boundary gravitational flux.

There is in principle some change of variables that writes this boundary integral in terms of variables in the original local theory.  Since the gravitational flux is a (gauge-invariant) bosonic observable at the boundary of the gravity theory, we assume that the result  in the kinematically local theory is again the integral of a bosonic gauge-invariant operator supported only on (or near) the boundary. Failure of this property to hold would mean that the two theories have radically different notions of bulk vs. boundary; we therefore refer to the above property as the assumption that the two theories have compatible notions of locality.  But having expressed the Hamiltonian in terms of boundary operators in the local theory, it must commute with all local observables in the interior. So interior local observables must be time-independent in the limit where the effective gravitational description applies; i.e., the local interior dynamics has become frozen.  We restate this conclusion as the following theorem.

\begin{theorem}
 Consider any limit where the effective description of a local theory is a
 gravitational theory with universal coupling to energy, the same notion of time evolution, and a compatible definition of locality.  In this limit all local observables away from the boundary become independent of time.
\end{theorem}

\section{Discussion}
\label{disc}

We have seen that all bulk dynamics must freeze out in any limit where a kinematically-local theory develops an effective gravitational description (and where this gravitational field couples universally to energy, maintains the same notion of time evolution, and contains a compatible definition of locality).  We emphasize that only the kinematics need be local for this conclusion to hold.  While our definition of kinematic locality requires Hamiltonian evolution, the Hamiltonian may contain instantaneous long-range interactions.

As remarked in the introduction, there is no inherent contradiction in this freeze out on its own.  After all, gapped theories are quite common.  But an interesting effective gravitational description should remain non-trivial in the bulk of the spacetime,
which then requires that its notion of the bulk/boundary distinction be rather different than that of the original kinematically-local theory.  This constitutes a certain non-locality intrinsic to the process of emergence -- beyond any non-locality already present in the original dynamics -- and is similar to what occurs in string theoretic gauge/gravity duality (e.g. \cite{Banks:1996vh,Maldacena:1997re}).  Indeed, modulo anomalies we may imagine discarding the original bulk and obtaining the gravity theory directly from degrees of freedom at the boundary.

Since partial motivation for this work came from the (gravitational) Weinberg-Witten theorem \cite{Weinberg:1980kq}, one may recall that Weinberg-Witten has a useful analogue for U(1) vector fields.  The corresponding analogue of our result is far less interesting.  It states simply that all local operators remaining in the limit where the effective U(1) vector description applies must be uncharged.

Returning to the gravitational context, it is clear that the consequences of our theorem can be avoided by introducing a priori kinematic non-localities violating our assumptions.  The gauge/gravity dualities of string theory are examples of this strategy.  Indeed, any (Hamiltonian) quantum theory of gravity defined on a separable Hilbert space is completely equivalent to some local field theory -- and in fact to a quantum mechanical theory describing a single particle in one dimension -- via a sufficiently non-local map.  One simply uses the fact that all separable Hilbert spaces are isomorphic to transcribe the Hamiltonian to the Hilbert space of a single non-relativistic particle.  As a 0+1-dimensional field theory the result trivially satisfies definition II.  The dynamics are also local in time, though when written (perhaps only formally) in terms of the usual position and momentum operators the Hamiltonian need not bear any resemblance to standard energy functions of Newtonian mechanics.

Of course, the above construct requires one to first know the exact spectrum of the gravitational Hamiltonian.  This is tantamount to solving the theory.  And any construction which first the theory to be solved will be of very limited use. Allowing the map between theories to be arbitrary non-locality thus seems unproductive.  Again, stringy gauge/gravity duality represents a sort of happy medium with enough non-locality to evade our theorem and enough structure to remain useful.

One might also ask if gravity could be the effective description (via a more local change of variables) of a theory with some special type of kinematic non-locality over which one might hope to have more control.  Non-commutative gauge theories \cite{Connes:1994yd} are a natural first category to consider.  Since these theories lack local observables, there is no immediate direct transcription to this context of our theorem above.  But closely related reasoning indicates failure here as well.  In particular, recall that non-commutative gauge theories can be defined on compact spaces with translational symmetry (e.g., tori) where they continue to admit gauge-invariant observables with non-zero momentum \cite{Ishibashi:1999hs}.   Recall also that, like energy, momentum is a source for (other components of) the gravitational field and admits a similar Gauss-law expression as a boundary integral in standard gravitational theories.  This motivates a definition of ``universal coupling to momentum'' in analogy with our definition I above.  In theories with this property the total momentum must vanish on spatially compact manifolds, and operators with non-zero momentum cannot be gauge invariant.  But restricting the non-commutative theory to zero-momentum operators is comparable to freezing out bulk degrees of freedom in a local field theory \cite{Gross:2000ba}, so this approach seems similarly unproductive.

In closing, we remark that the momentum version of the argument in the above paragraph also constrains the emergence of Ho\v{r}ava-Lifshitz gravity, which couples universally to momentum but not to energy.  Again, this universal coupling is an intrinsically non-linear phenomenon.  Thus the linearized theory is free to appear in an effective description as found in \cite{Xu2006cond.mat..2443X,Xu2006PhRvB..74v4433X,Gu:2009jh,Xu:2010eg}.

\section*{Acknowledgements}
I thank Veronika Hubeny, Shiraz Minwalla, Hirosi Ooguri, Joe Polchinski, Mukund Rangamani, and Cenke Xu for motivation and useful conversations over many years.  I also thank Joe Polchinski and Cenke Xu for comments on earlier drafts. This work was supported in part by the National Science Foundation under Grant No PHY11-25915, by FQXi grant FRP3-1338, and by funds from the University of California.

\begingroup\raggedright\endgroup

\end{document}